# Probing interlayer van der Waals strengths of two-dimensional surfaces and defects, through STM tip-induced elastic deformations

N. Sarkar[1*], P.R. Bandaru,[1,3] R.C. Dynes[2,3*]

[1]Department of Mechanical Engineering, [2]Department of Physics,

[3]Program in Materials Science,

University of California San Diego, La Jolla, California 92093-0411, USA

E-mail: nirjharize@gmail.com, pbandaru@ucsd.edu and rdynes@ucsd.edu

**Abstract:**

A methodology to test the interlayer bonding strength of two-dimensional (2D) surfaces and associated one(1D)- and two(2D)- dimensional surface defects using scanning tunneling microscope (STM) tip-induced deformation, is demonstrated. Surface elastic deformation characteristics of soft 2D monatomic sheets of graphene and graphite in contrast to $NbSe_2$ indicates related association with the underlying local bonding configurations. Surface deformation of 2D graphitic moiré patterns reveal the inter-layer van der Waals (vdW) strength varying across its domains. These results help in the understanding of the comparable interlayer bonding strength of 1D grain boundary (GB) as well as the grains. Anomalous phenomena related to probing 2D materials at small gap distances as a function of strain are discussed.

Atomic scale modulations of 2D materials using scanning tunneling microscope (STM) tip-induced forces modifies their mechanical and electrical properties(1,2). A better understanding of related atomic deformation mechanisms is relevant to the emergent fields of twistronics(3) and straintronics(4). Other conventional methods of using strain as a degree of freedom has brought forth unexpected characteristics in 2D materials *e.g.,* superconductivity (5) and ferromagnetism (6) in bilayer graphene and chalcogenides induced through hetero strain (4), hydrostatic pressure, or substrate engineering (7). We show that atomic scale control through tip-induced deformations on the material surfaces provides an alternate methodology for atomic strain engineering using STM. Such tip-induced deformations have not much been explored for the novel manifestations of the interlayer separation degrees of freedom, e.g., as related to electronic flat band formations on moiré domains with strain (6) or induction of soliton-like behavior of domain walls (DW) in moiré heterostructures (8). As yet another example, it has been predicted (9,10) in bilayer graphene that tuning the interlayer separation can control the band gap and effective mass of charge carriers. In this paper, we additionally illustrate a methodology to estimate interlayer bonding strength of topological defects in graphite and other 2D material surfaces via controlled elastic deformation of atomic bonds through STM electrical tunneling induced forces (11–13). The type and magnitude of the strain in varied 2D systems was observed to be a function of the tip-sample interaction and local atomic bonding configuration. The ease of exfoliation of the bulk layered materials and resultant surface features are related to their structural anisotropy. The constituted exfoliated heterostructures (3) would be dependent on issues related to local bonding, anisotropy, etc. and need to be investigated for insights and potential technological implications.

We show that STM and associated spectroscopy can yield significant insights into such strain-based effects. For instance, a large (/small) tip-sample gap as monitored through a small (/large) tunneling current will induce small (/large) deformation of the 2D layer: Fig. **1(a).** These layered 2D structures exhibit softer elastic properties that differ significantly from rigid 3D materials. The unique capability of STM to probe the influence of various tip induced forces considering conflation of various electronic and mechanical tip-sample interactions(14,15) could be harnessed, e.g., in one- and two-atom based layers, through graphitic sheets and $NbSe_2$, respectively.

Moreover, while highly oriented pyrolytic graphite (HoPG) has been extensively used in STM studies as a calibration standard (16,17) due to its regular step height arising from atomic flatness and surface cleanliness. However, previous STM measurements have indicated some confusion on its atomic amplitudes (16,17). For instance, while the measured atomic amplitudes on graphite surfaces are expected to be 0.2Å from Local Density of States (LDOS) calculations, abnormally high atomic amplitudes up to 24Å were reported in experiment (11,12). While the oscillations follow the modeled LDOS based contours (15), earlier tip geometry based electronic models (18–20) were inadequate to quantitatively account for anomalous high atomic amplitudes. Prevailing dirt-based model (12) where possible surface contaminants act as compressible springs between tip and sample causing anomaly, also seem contradictory to our observations. Such high amplitudes have been tentatively attributed to the tip-surface interactions which may induce local elastic deformation (11,13) of graphite surfaces particularly at small gap distances based on an interatomic potential model (11,21). While Atomic Force Microscopy (AFM) can be used to measure such tip-sample forces, in our experiments the focus has been on the atomic deformation of 2D material surfaces. The estimation of this atomic corrugation requires atomic resolution which is not easy in AFM but very feasible in STM and it also allows a finer control in tip height adjustments through tunneling currents. We perfomed an extensive experimental investigation of the deformation of 2D material surfaces and surface features with varied elastic properties to comparatively probe their interlayer vdW strength.

## 2. Results and Discussion

The anomalous atomic amplitudes are investigated on the surfaces of graphite, NbSe2 and single layer graphene (SLG) as well as across surface features such as monatomic step heights of NbSe$_2$ (in figure **1(b)**), monoatomic graphite step (in figure **1(c)**), 1D graphite grain boundary (in figure **1(d)**), 2D moiré pattern (in figure **1(e)**) as well as 1D wrinkle on graphite surface (in figure **1(f)**). The measured atomic corrugation amplitudes of graphitic surfaces, NbSe$_2$ and SLG is compared as a function of $I_{tip}$ (at a constant $V_{sample}$): figure **2(a)**, $V_{sample}$ (at a constant $I_{tip}$): figure **2(b)** and $R_{gap}$ (at a constant $I_{tip}$ and $V_{sample}$): figure **2(c)**. The same tip was used for all three materials over a fixed range of $I_{tip}$ and $V_{sample}$. An enhancement of atomic amplitudes on graphite and NbSe$_2$ surface is observed while the step height is relatively constant. Graphite sheets are observed to be one-layer thick separated by ~ 3.2 Å while a NbSe$_2$ surface sheet would be three-atoms (/layers) thick at ~12.5 Å. Consequently, the step height-based standard in graphite samples is reconfirmed as an accurate height calibration feature in the STM. More details on the topography across monatomic step heights of graphite and NbSe$_2$ as a function of tunneling gap distances is indicated in *Supplementary Section S1*.

Previously, the enhanced amplitudes with increasing $I_{tip}$, as seen in figure **2(a)**, were ascribed to electronic based corrugation(18,22) from varying LDOS related features in the STM images of semiconductor surfaces(23). Such electronic variation derived corrugations were also predicted(22) to be enhanced with reduction in $V_{sample}$ that would flatten out with increasing voltage, as experimentally confirmed in figure **2(b)**. However, according to these electronic models(18,22), the tunneling tip has to be closer to the surface for the imaging conditions when $I_{tip}$ > 10 nA (at $V_{sample}$ of 100 mV) in figure **2(a)** and when V$_{sample}$ < 100 mV (at $I_{tip}$ of 10 nA) in figure **2(b)**. Tip displacements of a few angstroms would then imply(11,18,22) physical contact of tip with surface, indicating(11,12) that the actual tip displacement cannot be solely a matter of the electronically induced corrugation. Alternately, it was proposed(11,13) that the tip elastically deforms the surface to produce corrugation which results from a force acting locally over atomic distances. Such atomic forces would be stronger at smaller gap distances and thus corrugation amplitudes would be function of tunneling gap resistance: $R_{gap}$ (= $V_{sample}$ / $I_{tip}$) which diminishes with the decrease in gap distance.

The amplitude enhancements in Figs. **2(a)** and **2(b)** can be best understood when the $I_{tip}$ increment and $V_{sample}$ reduction is seen as consequential reduction in $R_{gap}$ (shown by green arrow on top of figure 2 plots). The amplitude measurements of all three surfaces in Figs. **2(a)** and **2(b)** are then plotted in Figure **2(c)** as a function of $R_{gap}$ as shown by solid line and dashed line, respectively. Previous work showed (12) that the graphite atomic amplitudes enhanced at $R_{gap}$ < ~10 MΩ. as also experimentally observed in figure **2(c)** for not only graphite surface but also for NbSe$_2$ and SLG in our measurements. The onset of tip-induced atomic corrugation under ~10 MΩ appears to be not dependent on the type of material surfaces and could be indicative of a corresponding gap distance close to interatomic spacing where the interatomic tip-induced forces are strong enough to distort the 2D material surface.

The atomic corrugations via individual deformation of atomic layers by tip-induced forces can be understood through the comparison of the topography at two different tip-sample $I_{tip}$, *i.e.,* at 3nA and 30 nA, or $R_{gap}$ at 33 MΩ and 3.3 MΩ, respectively as shown in figure **3(a)**. The extrema in the topography is correlated to atomic deformation. For instance, the local maxima (/minima) corresponds to the outward (/inward) deformations of atoms. Closer (/farther) gap distances at larger (/smaller) currents corresponding to larger (/smaller) amplitude indicate atomic corrugation over a constant step height. A rationale, from elastic theory (11), suggests that inward deformation of the surface would be opposed by a strongly repulsive restoring force from the bulk in contrast to the outward deformation opposed by weaker attractive restoring forces. At small gap distances where corrugation is observed under $R_{gap}$ of ~10 MΩ, the tip could be acting against the steeper repulsive interatomic potential and deforming the surface. The typical interatomic potential [$V(r)$] profiles in figure **3(b)**, *e.g.,* the Morse(11) potential: $V(r) \sim \left[1 - e^{-a(r-r_o)^2}\right]$, with a as the potential profile width, $r$ as the distance between the atoms, and $r_o$ as the equilibrium distance, are asymmetric implying that the repulsive force is stronger compared to the attractive force. This suggested the inward deformation of atoms to be suppressed compared to the outward deformation as also evident from the fluctuations in figure **3(b)**. Local maxima appear to fluctuate more than the local minima. These fluctuations can also be represented in terms of standard deviation (SD) where it was observed that SD of the local minima is smaller compared to that of the maxima, as a function of the $I_{tip}$ as indicated in figure **3(c)**.

Generally, distorted surfaces would have different interatomic potential profiles than flatter non-distorted surfaces and thus consequently such interatomic potential profiles would be a function of deformation given the large standard deviation in the outward deformation.

The magnitude of deformation of all with these materials varies with their elastic surface properties and seem confirmative of the elastic deformation theory. In contrast to graphite, a smaller amplitude enhancement was seen in NbSe$_2$ in the range of 1.5Å to 2.5Å and SLG – in the range of 2Å to 2.5Å. It has been suggested (11,13–15) that the attractive (/compressive) forces from the tip cause an outward (/inward) deformation of the surface which would be manifested through elastic shear along the basal plane. Consequently, the extent of restoring shear forces, subsequent to tip-induced atomic deformation is larger in magnitude for $NbSe_2$ with heavier atoms, compared to a thinner monatomic sheet like graphite. Hence, the extent of deformation and atomic amplitudes would be smaller in $NbSe_2$ and larger in graphite, given similar tip-induced forces: *See supplementary section S2*. The measured deformations in Figure **2** are repeatable and thus within the elastic limit. However, plastic deformation using STM tip-induced forces has also been observed in our STM studies at very small gap distances, for example, tearing, folding and unfolding of graphite atomic layers at the step edges (24,25) which can be controlled for better manipulation of atomic layer sheets. At such a small tip to sample spacing, there is a risk of crashing the tip which changes the tip apex geometry and induces unusual deformation amplitudes.

A tip-surface interaction with the topmost layer of graphite as well as the SLG surface would be expected to yield comparable deformation. However, the SLG placed on an atomically smooth mica substrate, shows (See figure **2**) much lower atomic deformation amplitude compared to graphite. This suggests that the recorded anomalous corrugations of graphite surface are not just a surface interaction but involve bonding configurations from the bulk. The vdW bonding between the layers in graphite is distinctly weaker than the stronger Coulomb bonding of the SLG to the mica(26). Indeed, the extent of deformation for $NbSe_2$ and SLG is found to be similar, in contrast to graphite, suggesting that the SLG-mica adhesion may be comparable in magnitude to the vdW bonding between the $NbSe_2$ layers. Consequently, based on such results, tip-sample mediated atomic-scale deformation may be used for comparing the relative bonding strength of surfaces and constituent defects.

The electronic properties of stacked 2D materials are expected to depend on the nature of the interactions between the layers and their separation. Tip-induced deformation allows to compare the interlayer strength of such materials but can also be used to modulate the interlayer separation (8). A possible methodology has been demonstrated to test the bonding strength of 1D graphite defects, such as: grain boundary and wrinkles and 2D moiré patterns. The latter are suitable platforms to investigate the vdW bonding strength as a function of deformation, as the layers are slightly out of registry with respect to each other, yielding a long wavelength periodic pattern of inter-layer vdW/coupling strengths with varying topography and stackings (27,28). It is well known (27,28), for instance, that a relative twist between two sheets of SLG creates a moiré pattern with two kinds of domains broadly: (i) AA or BB type stacking,

i.e, the A or B atoms are directly on top of each other, or (ii) Bernal type: AB or BA stacking, i.e, the A (/B) atom occupy the hollows of the lower layer hexagons. The Bernal stacking is favored from an energetic stability point of view while the AA/BB stacking is unfavorable.

We demonstrate that the tip-induced surface deformation across the moiré pattern reveals local vdW bonding strength variability (14) as well as interlayer interactions (27,28). An instance of a pattern formed by a 4.2° twist angle in the upper layers of graphite is shown in figure **4(a).** AA domains are the bright/"hilly" regions while the AB domains are the dark/"valley" regions. The lines connecting the consecutive AA regions are termed domain walls (DWs). Scanning the tip along AA → AB → AA: *white* dotted line in figure **4(b)** at various $R_{gap}$ shows that the AA sites have deformed outward the most while the AB sites have deformed the least. The surface deformation amplitudes of these domains are estimated from figure **4(b)** and compared in figure **4(c)** as a function of $R_{gap}$. At a reduced tip-sample distance, there is considerable outward deformation/*bulging* out of the AA regions by ~11 Å owing to unstable stacking whereas the AB regions are much less deformed, *i.e.,* at ~3 Å given its stable stacking configurations (27,28). Thus, the tip-induced deformation methodology would be used to show that the weakly vdW bonded AA domains can be deformed more easily than the AB domains that are strongly bonded to the underlying layer, while DW shows intermediate response with $R_{gap}$. The relative deformation of the various moiré domains, with respective to the topography, reveals their inter-layer vdW strength. Such aspects can be utilized to probe the inter-layer bonding strength of grain boundaries and wrinkles, as well.

In yet another example, grain boundaries (GB) on graphite comprised of two grains oriented at an angle are shown in figure **4(d)**. The GB sites are buckled out of plane due to local strains induced by non-hexagonal dislocation cores (29). It would be nominally expected that such confined and buckled domains are weakly bonded to their underlying layer and may be deformed further with tip-induced forces. However, we observe that such GB sites are as strongly bonded as the Bernal stacked graphite. Scanning the tip across the GB along the *white* dotted line in figure **4(d)** indicates the extent of buckling at various gap distances in figure **4(e)**. The buckling amplitude of the GB is measured as a function of $R_{gap}$ in figure **4(c)**. At a reduced tip-sample distance, there is an outward deformation at the GB site of ~2 Å. Such deformation amplitude is similar to the AB site as compared in figure **4(c)** which indicates that such a buckled GB is comparable to Bernal stacked graphite. Similarly, deformation amplitudes of wrinkles have also been estimated comparatively in figure **4(c)** and appear to be as strongly bonded to the underlying layer as the Bernal stacked graphite. The deformation amplitudes of moiré domains and other defects in figure **4(c)** gradually appears to increase under ~10 MΩ, consistent with the deformation of the studied three surfaces. The demonstrated atomic force spectroscopy and imaging is not just restricted to studying surface elastic features and interlayer bonding strength but may also be useful for probing in-situ electronic modulations (1) with deformations as well as for analyzing simple adsorbates and soft biological materials.

## 3. Conclusion

The interplay between inter-layer coupling as well as the association with the substrate of typical one-atom and two-atom constituted 2D material systems, in response to electromechanical forces at the nanoscale, has been demonstrated. This was achieved through an analysis of the atomic scale corrugations and ascribed to the modulation of physico-chemical bonding attributes. The importance of atomic corrugation, with respect to $R_{gap}$ reveals the mechanical nature of the tip-induced elastic deformation model. The aspect of probing inter-layer bonding strength of pristine atomic surfaces along with 1D defects such as grain boundaries, wrinkles and unique topologies related to 2D moiré patterns is shown to be feasible through the controlled surface deformation. Such methodology opens possibilities in exploring in-situ the electronic and mechanical properties of 2D stacked structures useful for pressure sensitive devices and sensors.

## 4. Materials and methods

### 4.1 Materials, Exfoliation and Layer synthesis

Highly oriented pyrolytic graphite (HoPG) and niobium diselenide ($NbSe_2$) were freshly cleaved from crystals before loading into the STM chamber. The choice of HoPG with roughly micron sized grains (ZYH grade from Advanced Ceramics, Inc.) favors a dense occurrence of step heights as well as multiple moiré patterns with different twist angles. Single crystals of 2H-$NbSe_2$ with millimeter grain sizes were grown by chemical vapor deposition at AT&T Bell labs. Single layer graphene (SLG) was grown by chemical vapor deposition (CVD) with methane precursor on copper foil and subsequently PMMA polymer-assisted wet transfer of graphene was done onto atomically flat cleaved mica substrates. After dissolving the polymer and thermally evaporating Ti/Au contacts, the graphene/mica stack was baked in an oven at 250°C for a week to dehydrate and degas before loading into the STM.

### 4.2 Electrical Measurements

All imaging was done using a custom-built tunneling STM interfaced with a RHK controller at room temperature and atmospheric pressure. All topography measurements were performed in standard constant current mode at ~0.5 Hz scanning frequency using a mechanically snipped Pt/Ir tip.

### 4.3 Calibration and Data analysis

Topographic heights of all images were calibrated using the constant step height measured on graphite and $NbSe_2$ monatomic steps that remains constant irrespective of tunneling conditions. The same tip was used to measure the average atomic heights of graphite, $NbSe_2$ and SLG. Step heights are measured by the difference in mean of the two modulations on either side of the step. Atomic heights are averaged over 20 peak-to-peak amplitudes.

### Data availability statement

The experimental data and its analysis in the paper and/or in the supplementary information is sufficient to support our conclusions. Additional data can be made available on request.

### Acknowledgements

This work was supported by AFOSR Grant (FA9550-15-1-0218) and Army Research Office (AROW911NF-21-1-0041). The authors thank Michael Rezin for the technical assistance; Prof. Shane Cybart, Uday Sravan Goteti and Hidenori Yamada for useful discussions.

### ORCID iDs

Nirjhar Sarkar  https://orcid.org/0000-0002-0093-9523

Prabhakar .R. Bandaru  https://orcid.org/0000-0003-4497-9620

Robert .C. Dynes  https://orcid.org/0000-0001-6740-9677

Acknowledgements:

This work was supported by AFOSR Grant (FA9550-15-1-0218) and Army Research Office (AROW911NF-21-1-0041). The authors wish to thank Michael Rezin for the technical assistance; Prof. Shane Cybart, Uday Sravan Goteti and Hidenori Yamada for useful discussions.

Author contributions:

N. Sarkar did the experimental work and along with P.R. Bandaru and R.C.Dynes, wrote the paper. All analysis and discussion were under the supervision of R.C. Dynes and P.R. Bandaru.

Data availability:

The experimental data and its analysis in the paper and/or in the supplementary information is sufficient to support our conclusions. Additional data can be made available on request.

# Figure

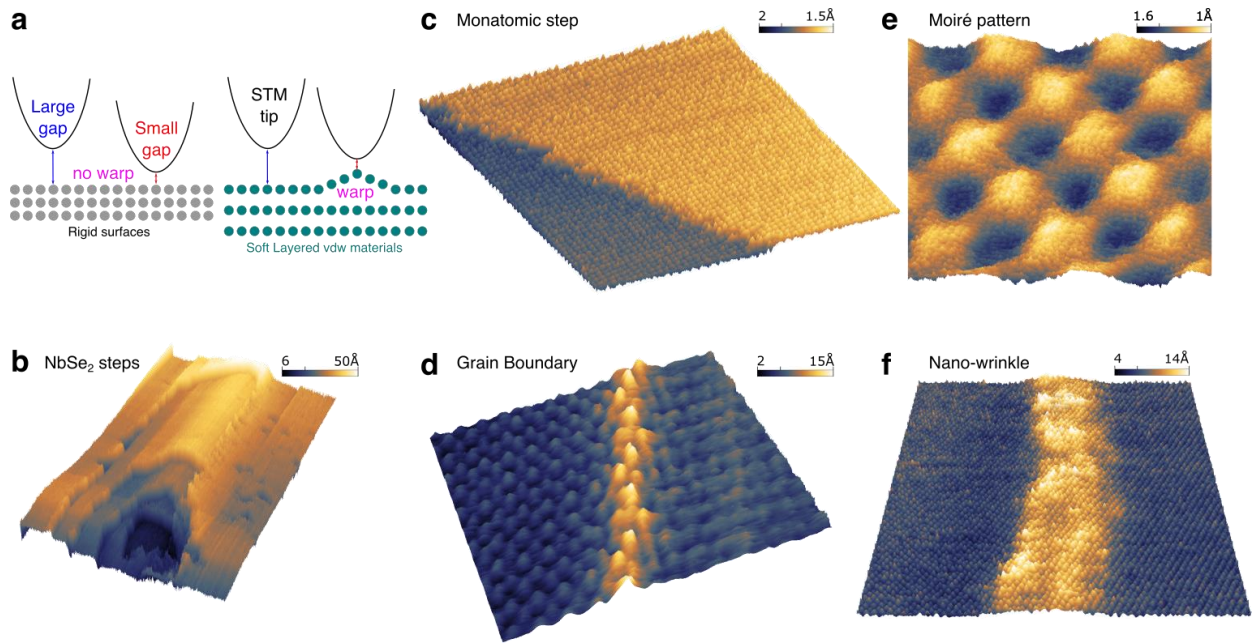

Figure 1: **Tip-induced atomic deformation of surfaces and features of graphite and NbSe$_2$.** **(a)** The expected response of *rigid* vs. *soft* 2D materials to scanning tunneling microscopy (STM) tip induced deformations at large (/small) tip-sample gaps monitored by higher (/small) tunneling currents ($I_{tip}$). The topographic images at an $I_{tip}$ of 30nA and 100mV $V_{sample}$ indicate **(b)** series of monatomic steps on NbSe$_2$ surface (230 nm x 230 nm) and **(c)** monatomic step on graphite surface (11.9 nm x 9.5 nm) **(d)** 1D graphite grain boundary (3.8 nm x 3.1 nm) with a relative grain orientation of 28° **(e)** Atomically resolved three-dimensional topography of a moiré pattern with 4.2° twist angle between top and bottom layer showing bright hills (AA atomic layer stacking) and dark valley (AB atomic layer stacking) domains joined by DW (Domain Walls bridges). **(f)** Nano-wrinkle on top graphite sheet (15 nm x 12 nm). The small dots across these images correspond to individual atoms. The scale indicates the measured height in Angstroms (Å).

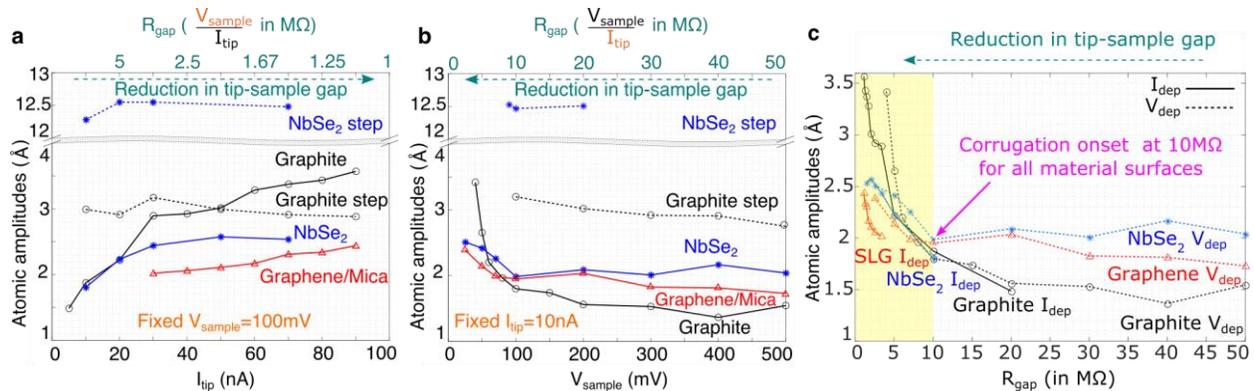

Figure 2: **The atomic corrugation amplitude as a function of tunneling current ($I_{tip}$), sample voltage ($V_{sample}$) and gap resistance ($R_{gap}$)**, An increase in the **(a)** $I_{tip}$, or equivalently a decrease in **(b)** $V_{sample}$, is related to a reduction in tip-sample gap, and yields higher amplitude of atomic oscillations. A larger increase was found on the graphite surface compared to NbSe$_2$ and single layer graphene (SLG). **(c)** The $R_{gap}$ was obtained through the ratio of $V_{sample}$ to $I_{tip}$ from **(a)** and **(b)** also labeled as $I_{dep}$ (shown by solid line) and $V_{dep}$ (shown by dashed line). Overlap of $I_{tip}$ and $V_{sample}$ dependence curves for a given material indicates $R_{gap}$ dependence as predominant dependent variable. Irrespective of the material, 10 MΩ appears to be the onset of tip-induced atomic corrugation.

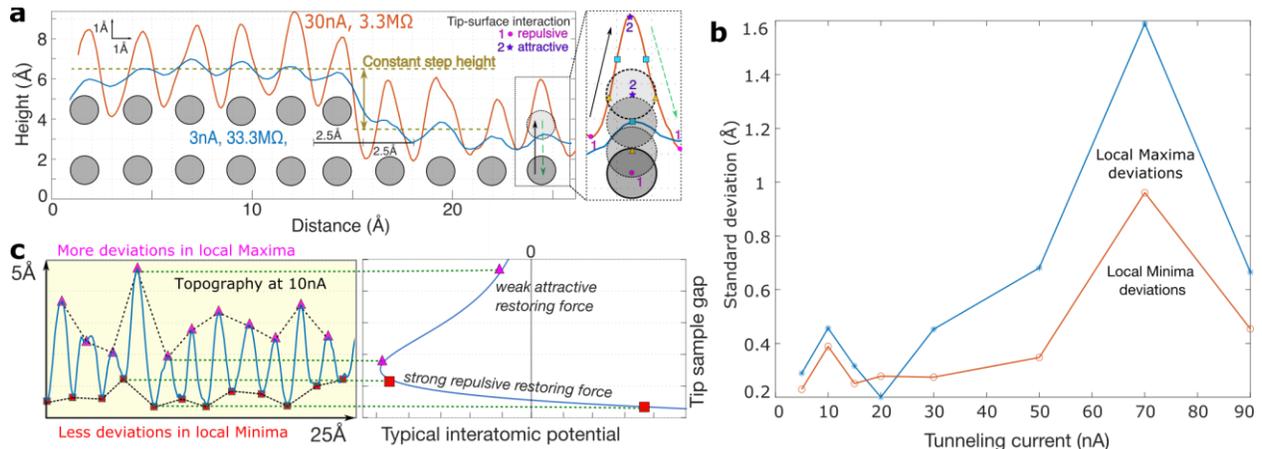

Figure 3: **Asymmetric features in atomic topography across a monatomic step**
**(a)** Comparing the topography at two different tip-sample distances, i.e., closer (/further) spacing corresponding to an $I_{tip}$ of 30 nA (/3nA) indicates a larger (/smaller) atomic corrugation amplitude. The right inset indicates a rationale for the higher corrugations through tip induced local outward deformation of an atom. **(b)** Standard deviation (SD) of the local maxima of the atomic deformation amplitudes is larger than that of the minima, as a function of the $I_{tip}$. Deformations are asymmetrical as shown in **(c)**. **(c)** The correlation of the atomic topography to typical interatomic potential curves. The weaker attractive forces between the tip and the surface layer (bounded by air) exhibit strong fluctuations in contrast to the much smaller variations from the stronger repulsive force from the bulk.

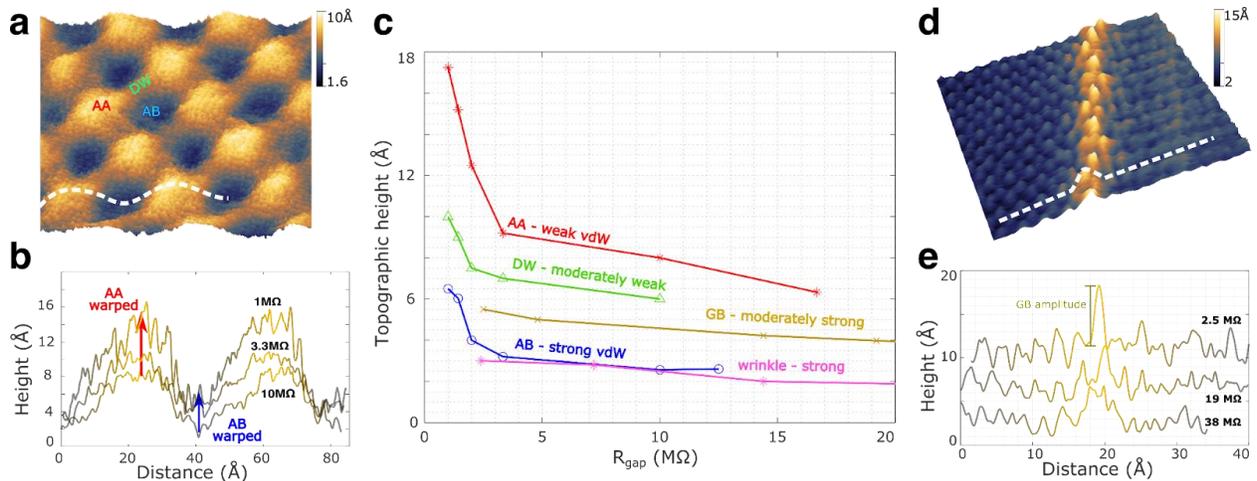

Figure 4: **Probing interlayer interaction strength of 2D moiré pattern and 1D grain boundary** **(a)** Three-dimensional atomically resolved topography of Moiré pattern of twist angle 4.2° consisting of hills (AA domains), valleys (AB domains) and bridges (Domain Walls) as labeled. A scan direction (hill → valley → hill: *white* dotted line is shown. Atomic amplitude variations and surface deformation along this line are exhibited at differing $R_{gap}$ in **(b).** A high (/low) surface deformation is observed in the AA (/AB) region, due to the weak (/strong) vDW-related bonding. **(c)** Three-dimensional atomically resolved topography of a GB with a relative grain orientation of 28° exhibiting the buckling of the GB site. **(d)** Topography variation only along the *white* dotted line in **(c)** at differing $R_{gap}$. **(e)** Deformation amplitudes of the moiré features and GB sites are estimated from **(b)** and **(d)** as a function of $R_{gap}$. A larger (/smaller) extent of deformation is seen for the AA (/AB) sites with an intermediate extent for the DWs and GBs. Similar deformation amplitudes are observed for nano-wrinkle and AB sites.